# Effect of manganese doping on the size effect of lead zirconate titanate thin films and the extrinsic nature of "dead layers"

X.J. Lou[*] and J. Wang

Department of Materials Science and Engineering, National University of Singapore, Singapore 117574, Singapore

**Abstract:**

We have investigated the size effect in lead zirconate titanate (PZT) thin films with a range of manganese (Mn) doping concentrations. We found that the size effect in the conventional Pt/PZT/Pt thin-film capacitors could be systematically reduced and almost completely eliminated by increasing Mn doping concentration. The interfacial layer at the electrode-film interface appears to disappear almost entirely for the PZT films with ~2% Mn doping levels, confirmed by the fits using the conventional "in-series capacitor" model. Our work indicates that the size effect in ferroelectrics is *extrinsic* in nature, supporting the work by Saad *et al*. Other implications of our results have also been discussed. By comparing a variety of experimental studies in the literature we propose a scenario that the "dead layer" between PZT (or barium strontium titanate, BST) and metal electrodes such as Pt and Au might have a *defective* pyrochlore/fluorite structure (possibly with a small portion of ferroelectric perovskite phase).

---

[*] Electronic mail: mselx@nus.edu.sg

**I. Introduction:**

Size effects, defined as the systematic degradation of functional properties with shrinking geometry, have long been one of the most serious problems hindering the miniaturization and application of ferroelectric components, along with imprint,[1] electrical breakdown[2] and polarization fatigue.[3] For ferroelectric materials, size reduction has significant impacts on their functional behaviors.[4,5] For instance, it has been reported that the remanent polarization $P_r$ decreases,[6] the coercive field $E_c$ increases[7] for the samples with reduced thicknesses. Furthermore, the measured dielectric constant (also called "relative permittivity") $\varepsilon(d)$ collapses with decreasing film thickness $d$.[4] The decrease in dielectric constant of ferroelectric thin films can be effectively modeled by assuming the existence of a low-$\varepsilon$ layer at the film/electrode interface.[8] The interfacial layer behaves like a parasitic capacitor connected in series with the ferroelectric bulk. The "in-series capacitor" model formed accounts for the collapse of dielectric constant in thin enough films, where the interfacial capacitance becomes a more important factor influencing the overall dielectric response. Although the interfacial-layer concept has been shown to be successful in explaining many of the experimental data in the literature, an experimentally consistent explanation for the nature of such layers is still lacking. Various models proposed in the past include a defect/space-charge layer with low $\varepsilon$ at the electrode-ferroelectric interface,[9] termination of chemical bonds at the interface,[10] interdiffusion of elements,[11] chemically different phases/layers,[12] changes in spontaneous polarization and polarizability of surface layers,[13-15] polarization reduction at the film surface due to an increase in the depolarization field as film thickness decreases,[16] Schottky barrier formation and the resultant surface depletion layer,[17] finite electronic screening length in metallic electrodes,[18] strain[19]/strain-gradient coupling[20] at the ferroelectric-electrode interface and so forth. However, none of the above is fully consistent with the extensive body of experimental observations published previously, and a complete understanding of the origin of size effects in ferroelectric thin-film capacitors has not been achieved so far.

In addition, the use of dopants to improve the electrical properties of a ferroelectric capacitor has been extensively investigated in the past. For example, doping could lead to enhanced fatigue resistance,[21,22] lowered leakage, smaller $E_c$,[22] and so on. Surprisingly, the effects of doping on the issue of size effects have been rarely studied, and the current knowledge about this issue is very poor. In particular, the effects of doping species and doping concentrations on the interfacial properties of ferroelectric thin films of a range of thicknesses remain largely unknown.

In this paper, we have investigated the effect of manganese (Mn) doping on the size effect of the conventional Pt/PZT/Pt thin-film capacitor [PZT denotes Pb(Zr/Ti)O$_3$]. The implications of our present work have also been discussed.

**II. Experiment:**

The Pb[(Zr$_{0.3(1-x)}$/Ti$_{0.7(1-x)}$)Mn$_x$]O$_3$ (PMZT) thin films with x=0, 0.2%, 0.5%, 1%, 2%, 3% of various thicknesses used in the present work were fabricated on Pt/TiO$_x$/SiO$_2$/Si substrates using sol-gel spin coating method (the samples are therefore coded as PZT, PMZT0.2%, PMZT0.5%, PMZT1.0%, PMZT2.0%, PMZT3.0% hereafter). 10% Pb-excess solutions of 0.4 M was used to prepare the P(M)ZT films to compensate Pb loss during annealing. After a desired number of layers were coated, the P(M)ZT films were finally annealed at 650 ºC for 30 min in air in a quartz tube furnace (Carbolyte) to achieve the wanted phase. The preparation procedures are as the same as those reported elsewhere.[23] Phase identification and growth orientation of the P(M)ZT films were characterized using X-Ray Diffraction (XRD, D8 Advanced Diffractometer System, Bruker). Film textures were measured using Field Emission Scanning Electron Microscopy (FE-SEM, XL30 FEG Philips). For electrical measurements, the top electrodes (Pt squares of ~100 x 100 μm$^2$) were then deposited by sputtering on the films via transmission electron microscopy (TEM) grids. Ferroelectric and dielectric measurements were conducted using a Radiant ferroelectric test system and an impedance analyzer (Solartron SI 1260), respectively.

### III. Results and discussion:

Fig 1 shows the XRD patterns of the ~170 nm undoped and Mn doped PZT films. It can be seen that the P(M)ZT thin films have a perovskite structure and are polycrystalline in nature. The PZT film is highly [111] textured, while the preferred orientation changes to [100] direction for the PMZT films with Mn ≥ 1.0%. Fig 2(a-d) show the FE-SEM micrographs of the 54 nm PZT film, the 481 nm PZT film, the 54 nm PMZT3.0% film and the 474 nm PMZT3.0% film. From the SEM studies of all these samples we conclude that thicker films or films with higher Mn doping levels generally contain larger grains, and vice versa. Cross-section SEM studies demonstrate that the films have a columnar microstructure [see Fig 2(e)].

Fig 3 displays the hysteresis loops measured at 1 kHz for the P(M)ZT films of ~170 nm in thickness with various Mn doping levels at room temperature. We see that the remanent polarization $P_r$ generally decreases as the Mn doping level increases: e.g., $P_r$ for PZT is 27.6 $\mu C/cm^2$ while it becomes 17 $\mu C/cm^2$ for PMZT3.0%. Our results are in good agreement with the work by Victor *et al.*,[24] who also found that the hysteresis loop was suppressed with the increase in Mn doping concentration for their $Pb_{1.05}(Zr_{0.53}/Ti_{0.47})O_3$ thin films.

The dielectric behaviors of the ~170 nm undoped and Mn doped PZT films are shown in Fig 4. The capacitance and loss measurements were conducted with zero bias voltage; 5 to 7 electrodes were measured for each sample to ensure that the measured values are indeed representative of the real value for each film. The dielectric constant value for each electrode at a specific frequency (e.g. $10^5$ Hz) was then evaluated by its own capacitance value, it own electrode/capacitor area measured using optical microscope [note that the electrode area could be quite different even for those deposited on the same film via TEM grids by dc sputtering due to the spreading effect of electrode material underneath the grid; using the same electrode area (e.g.

100x100 μm$^2$) for all the electrodes or even for all the films is therefore no good and may give rise to a wrong (normally overestimated) value for the derived dielectric constant], and the film thickness measured using Cross-section SEM. To ensure that all the samples with different thicknesses were stimulated by the ac field of similar magnitude (~2 kV/cm) during dielectric measurements, the oscillator levels were adjusted accordingly, i.e, 0.01 V for the ~52 nm samples, 0.05 V for the ~270 nm samples, and 0.09 V for the ~470 nm samples.

Fig 5(a) and Fig 5(b) show the room-temperature dielectric constant and loss tangent at $10^5$ Hz as a function of film thickness for the P(M)ZT films with various Mn doping contents. It can be seen that the dielectric constant of the undoped PZT thin films decreases significantly as film thickness is reduced. That is, a *strong* size effect is shown for the pure PZT films, which is consistent with the observations in the literature for both PZT (Ref [8]) and BST [Ref [10], BST denotes (Ba/Sr)TiO$_3$] with Pt or Au metal electrodes. However, as the Mn doping concentration increases to 0.2% and 0.5%, the size effect in the PMZT samples becomes significantly reduced [see Fig 5(a)]. The dielectric constant vs. film thickness plots become flattened when Mn doping level increases up to 1% and 2%, indicating that the size effect is gradually eliminated in these samples [see Fig 5(b). We also noticed that there is some variation in the dielectric-constant values as the thickness changes, i.e. the lines are not smooth. That is because our films, as many of those studied by other academic researchers and published in the literature, were made in the common laboratory in ambient atmosphere, not in the industrial clean room; smooth lines of dielectric constant vs thickness are not expected in general from these samples. Non-smooth, sometimes quite large, variation of *ε* with *d* has also be reported by Amanuma *et al*. (see Fig 4 in Ref [8]), by Lee *et al*. (Fig 1 and Fig 7a in Ref [25]), and by Sinnamon *et al*. (Fig 3 in Ref [26]). So this phenomenon is indeed common for the lab-made samples. That is also the reason why we made nine samples of different thicknesses for each doping level in order to ensure that our results are reliable.]. The precursor with Mn doping content equal to 4% was not stable and it formed precipitates within only one hour. Therefore, the PMZT films with Mn doping concentration equal to or more than 4% were not obtained.

The data shown in Fig 5 can be analyzed using the conventional "in-series capacitor" model. By assuming $d_i \ll d$ (or $\varepsilon_f \gg \varepsilon_i$, $d_i$ and $\varepsilon_i$ are the thickness and relative permittivity of the interfacial layer, respectively. $\varepsilon_f$ is the relative permittivity of the bulk ferroelectric),[8] this model predicts:

$$\frac{1}{C} = \frac{d}{\varepsilon_f \varepsilon_0 S} + \frac{d_i}{\varepsilon_i \varepsilon_0 S} \quad (1)$$

where $C$ is the measured capacitance, $S$ is the capacitor area and $\varepsilon_0$ is vacuum permittivity. Defining $C \equiv \frac{\varepsilon \varepsilon_0 S}{d}$, we have:

$$\frac{d}{\varepsilon} = \frac{d}{\varepsilon_f} + \frac{d_i}{\varepsilon_i} \quad (2)$$

where $\varepsilon$ is the measured dielectric constant of the capacitor. From Eq (1) and (2), one can see that the interfacial-layer model predicts a linear relationship between $d/\varepsilon$ (or $1/C$) and $d$ with the slope of $1/\varepsilon_f$ and y-axis intercept of $d_i/\varepsilon_i$.

The $d/\varepsilon$ vs. $d$ plots are shown in Fig 6(a) and Fig 6(b), respectively, for the P(M)ZT films with different levels of Mn doping along with the linear fits according to the interfacial-layer model [Eq (2)]. The good fit of these data by a linear function as shown in Fig 6(a) and Fig 6(b) indicates that the interfacial-layer model is indeed valid for the P(M)ZT films. The fitted results for $\varepsilon_f$ and $d_i/\varepsilon_i$ as a function of Mn doping levels are shown in Fig 6(c). From Fig 6(c), one can see that the $d_i/\varepsilon_i$ value becomes significantly reduced from 0.18 to 0.034 as the Mn doping concentration increases from 0 to 1.0%; this value approaches zero (i.e., ~0.0067) for PMZT2% and slightly increases for PMZT3%. The small and near-zero values of $d_i/\varepsilon_i$ indicate that either $d_i$ is extremely small or $\varepsilon_i$ is enormously large (i.e., a bulk-like value and close to $\varepsilon_f$). In both cases, a value of $d_i/\varepsilon_i$~0 indicates the close-to-disappearance of the interfacial layer. Interestingly, $\varepsilon_f$ also decreases as Mn doping concentration increases [Fig 6(c)].

Although the reason why PZT thin films doped with only ~2% Mn are almost free of size effect is still unclear at the present stage of research and warrants further investigations, some definite conclusions can still be drawn. Recall the explanations and scenarios proposed for the nature of the interfacial layer at the beginning of the paper. Since the electrodes (i.e., platinum) used in the present work for both PZT and PMZT are the same, the PMZT2.0% films should also show strong size effect if the finite electronic screening length in metal electrodes is the main cause of the size effect in these films. However, that is not the case. Therefore, the scenario invoking the finite electronic screening length in metal electrodes does not seem to be the main reason for the collapse in dielectric constant in thinner PZT films, at least for the thicknesses that are investigated here and of great interest in the literature, i.e., from ~50 nm to ~470 nm.

More importantly, our results suggest that the interfacial-layer effect in ferroelectrics is *not* intrinsic but *extrinsic* in origin, which is fundamentally different from the viewpoints in some papers in the literature where the interfacial layer was believed to be *intrinsic*. [by "extrinsic" (or "intrinsic"), we mean this effect could (or could not) be eliminated by improving the film quality using methods such as doping or changing electrode materials and so on; for instance, we could get rid of the interfacial layers *almost* entirely in Pt/PZT/Pt by ~2% Mn doping in the present work; the size effect in PZT and BST films could also be eliminated by adopting conductive oxide electrodes instead of metal electrodes (see Ref [27] and the references therein). Note that whether the size effect in ferroelectric thin films or the origin of the interfacial layer is *intrinsic* or *extrinsic* in nature is one of the most important questions that need to be answered]. Therefore, the models claiming that the collapse in dielectric constant in ferroelectric thin films is due to the *unavoidable intrinsic* change in spontaneous polarization or polarizability of electrode-film interfacial layers with decreasing film thickness might need to be reconsidered. Our results support the works by Saad *et al*., by investigating thin-film single crystal materials who argued that the dielectric collapse in thinner ferroelectric films is neither a direct consequence of reduced size nor an outcome of unavoidable physics related to the ferroelectric-electrode boundary.[28,29]

It has been conventionally supposed and recently confirmed both experimentally[30] and theoretically[31] that $Mn^{2+}/Mn^{3+}$ substitute for $Zr^{4+}/Ti^{4+}$ ions at the *B* site of $ABO_3$ perovskite structure as acceptors. However, further analysis of our data is complicated by our poor understanding of the semiconducting nature of ferroelectric thin-films capacitors. For many years, whether ferroelectric thin films (e.g. PZT or BST) with metal (e.g. Pt or Au) or conductive oxide electrodes (e.g., $IrO_2$ or $SrRuO_3$) are fully or partially depleted has been a matter of debate (see. e.g., Chapter 5 of Ref [1]). Alternatively, the answer to this question may be strongly sample-dependent. If we assume that the virgin film is fully depleted, doping with acceptors like Mn ions introduces more oxygen vacancies and therefore gives rise to higher space-charge density $N_d$. Since $W_d \sim 1/\sqrt{N_d}$,[32] where $W_d$ is depletion width, we would expect that an increase in Mn doping level will lead to progressively reduced $W_d$, the appearance of an in-series capacitor structure (i.e., a partially depleted structure) and consequently size effects in the thin films. However, if we assume that the virgin/doped film is partially depleted, further doping with Mn ions will result in a continuously reduced $W_d$ and consequently diminishing size effects in the samples. Since the origin of the interfacial layer in the topic of size effects is still unclear as aforementioned, our present results should not be used to make any judgment regarding whether our P(M)ZT films are fully or only partially depleted at the moment.

The classical semiconductor physics and the associated fully/partially depletion scenario mentioned above treat the metal-semiconductor or semiconductor-semiconductor interface as a perfect and sharp (i.e., free of interfacial layers) boundary (with/without interface states) between two stoichiometric and diffusion-free substances.[32] However, this assumption may not be satisfied for ferroelectric thin-film capacitors annealed for crystallization at high temperatures (e.g., 650 ºC to 800 ºC). It is well known in the literature that a certain degree of diffusion of ferroelectric and electrode materials occurs at the interfaces of ferroelectric capacitors during annealing.[33-37] The work by Wang *et al*.[35] is particularly interesting in that they showed that the BST/$RuO_2$ interface is superior to the BST/Pt and BST/Ru interfaces and therefore results in a structure with thickness-independent dielectric constant and free of size effects. This work is consistent with the results summarized by Jin and Zhu *et al*.[27] The better performance of conductive oxide electrodes such as $RuO_2$ and $SrRuO_3$ in contact with PZT or BST has been generally attributed to their better chemical compatibilities with the ferroelectric bulk than metal electrodes such as Pt and Au in the literature.[35]

Furthermore, it was widely reported that significant Pb loss occurs during annealing Pb-containing ferroelectric materials such as PZT.[1,38] By using *in situ* X-Ray photoelectron spectroscopy (XPS), Chen and McIntyre were able to show that the PZT thin films reacts with Pt during deposition, forming a Pb-deficient and Ti/Zr-rich defective interfacial layer.[39] The perovskite structure of PZT is believed to tolerate only 2%-3% Pb deficiency while the pyrochlore (or fluorite) phase is expected to bear a very wide range of Pb excess or deficiency.[40,41] [Fact I; Note that the pyrochlore structure is intimately related to the defective fluorite phase. It is a superstructure of the ideal fluorite structure, $AO_2$, with the *A*-site and *B*-site cations ordered and one-eighth of the oxygen anions missing.[42]]. By conducting dielectric and current-voltage measurements on a Pt/BST/$YBa_2CuO_7$ capacitor, Chen *et al*.[43] estimated $d_i$ and $\varepsilon_i$ to be 2.8 nm and 42.6, respectively. This value ($\varepsilon_i$~42.6) is consistent with our previous estimation of $\varepsilon_i$~40,[44] based on the work by Larsen *et al*.,[45] who showed that $d_i/\varepsilon_i$=0.036~0.05 nm in the Pt/PZT/Pt thin-film capacitors, and the study by Lee *et al*., who obtained $d_i/\varepsilon_i$=0.048~0.096 nm in Pt/BST/Pt, depending on the annealing conditions.[10] (we obtained the $d_i/\varepsilon_i$ value of 0.0067-0.18 for the PZT thin films with various Mn doping levels in the present work; note that the larger the $d_i/\varepsilon_i$ value the thicker the interfacial layer would be). Interestingly, the dielectric constant of pyrochlores was indeed found to be <100 and ~50 (see Ref [46]) [Fact II]. By using micro-Raman microscopy, Lou *et al* showed that the electrically degraded phases in PZT and Sm-doped $Bi_4Ti_3O_{12}$ thin films consist mostly of a pyrochlore-like phase after breakdown[2] and polarization fatigue[47], caused by charge injection and local Joule heat. The progressive degradation in dielectric constant of ferroelectric capacitors during bipolar electrical fatigue was reported by Mihara *et al*.,[34] by Jiang *et al*.,[48] and by Wang *et al*[49] [Fact III]. The long-time annealing studies of Jiang *et al* show that the thermally degraded PZT capacitor after heat treatment at 350 ºC for 257 hours has $\varepsilon \sim 48$, suggesting the growth of the interfacial layers throughout the whole film [Fact IV]. Work by Basceri *et al*.[19] shows that $d_i/\varepsilon_i$ is approximately independent of temperature in Pt/BST/Pt capacitors (i.e., ferroelectrically dead), while Finstrom *et al.* showed that the interfacial capacitance densities (or $d_i/\varepsilon_i$) is *weakly* temperature dependent in $Pt/SrTiO_3/Pt$ films[50] (that is, the interfacial layer is only ferroelectrically inferior and still a little active). Parker *et al*.[5] also believed that the low-$\varepsilon$ interfacial layer is ferroelectrically active, at least for their Pt/BST/Pt films [Fact V].

All these facts [Fact I-V] point to a scenario that the interfacial layer may have a *defective* pyrochlore/fluorite-like structure (possibly with a small portion of ferroelectric perovskite phase) and has $\varepsilon <100$ and $\sim 50$ for the Pt(Au)/PZT/Pt(Au) structure and probably for Au(Pt)/BST/Au(Pt) as well. This hypothesis is consistent with the extrinsic nature of this layer discussed above. If this scenario is correct, the reason why the PMZT2% films are almost free of size effect may be explained by the improved stability of the perovskite structure against Pb loss and consequently against the formation of a defective pyrochlore-like structure during annealing.

## IV. Conclusions:

In summary, we have studied the size effect in P(M)ZT thin films with various levels of Mn doping. We found that the size effect of the prototype Pt/PZT/Pt structure could be tuned, reduced and almost completely eliminated by increasing Mn doping concentration. The interfacial layer appears to entirely disappear for PMZT2% as confirmed by the fits using the "in-series capacitor" model. Our results suggest that size effects in ferroelectrics are *extrinsic* in origin and support the conclusions drawn by Saad et al.[28,29] By comparing various experimental works in the past we arrive at a scenario that the interfacial layer (or "dead layer" called by some researchers) between Pt/Au and PZT might have a *defective* pyrochlore/fluorite structure (possibly with a small amount of ferroelectric perovskite phase). We hope that the present work could shed new light on the long-standing problem of size effects in ferroelectric thin-film capacitors.

**Acknowledgement:**

X.J.L would like to thank the LKY PDF established under the Lee Kuan Yew Endowment Fund for support. This work is supported also by National University of Singapore, and a MOE AcRF grant (R284-000-058-112).

**Figure Captions:**

Fig 1 (Color online) XRD patterns of the ~170 nm P(M)ZT films with a variety of Mn doping concentrations.

Fig 2 (Color online) FE-SEM micrographs of (a) the 54 nm PZT film, (b) the 481 nm PZT film, (c) the 54 nm PMZT3.0% film and (d) the 474 nm PMZT3.0% film; (e) a cross-section SEM showing the columnar structure of a PZT film.

Fig 3 (Color online) hysteresis loops for P(M)ZT of ~170 nm in thickness with various Mn doping levels.

Fig 4 (Color online) Dielectric behavior of P(M)ZT thin films of ~170 nm in thickness with various Mn doping concentrations.

Fig 5 (Color online) dielectric constant and tangent loss at $10^5$ Hz as a function of film thickness for (a) PZT, PMZT0.2% and PMZT0.5%, and for (b) PMZT1.0%, PMZT2.0% and PMZT3.0%

Fig 6 (Color online) $d/\varepsilon$ vs $d$ plots and the linear fits by the "in-series capacitor" model for (a) PZT, PMZT0.2% and PMZT0.5%, and for (b) PMZT1.0%, PMZT2.0% and PMZT3.0%. (c) shows the changes of $\varepsilon_f$ and $d_i/\varepsilon_i$ obtained from the linear fits in (a) and (b) as a function of Mn doping level.

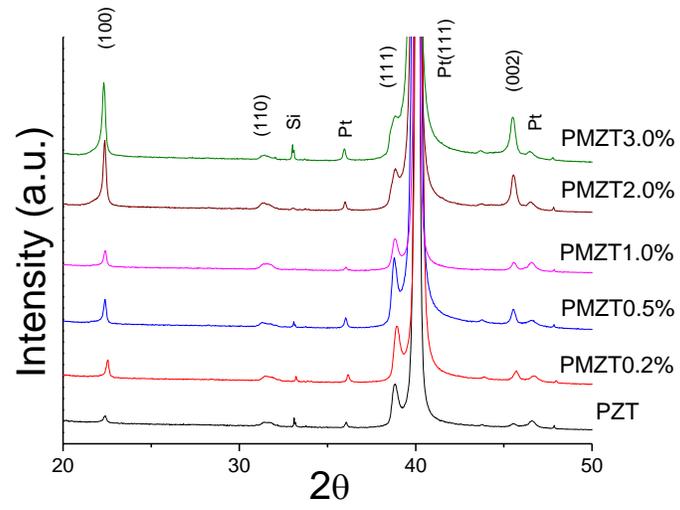

Fig 1

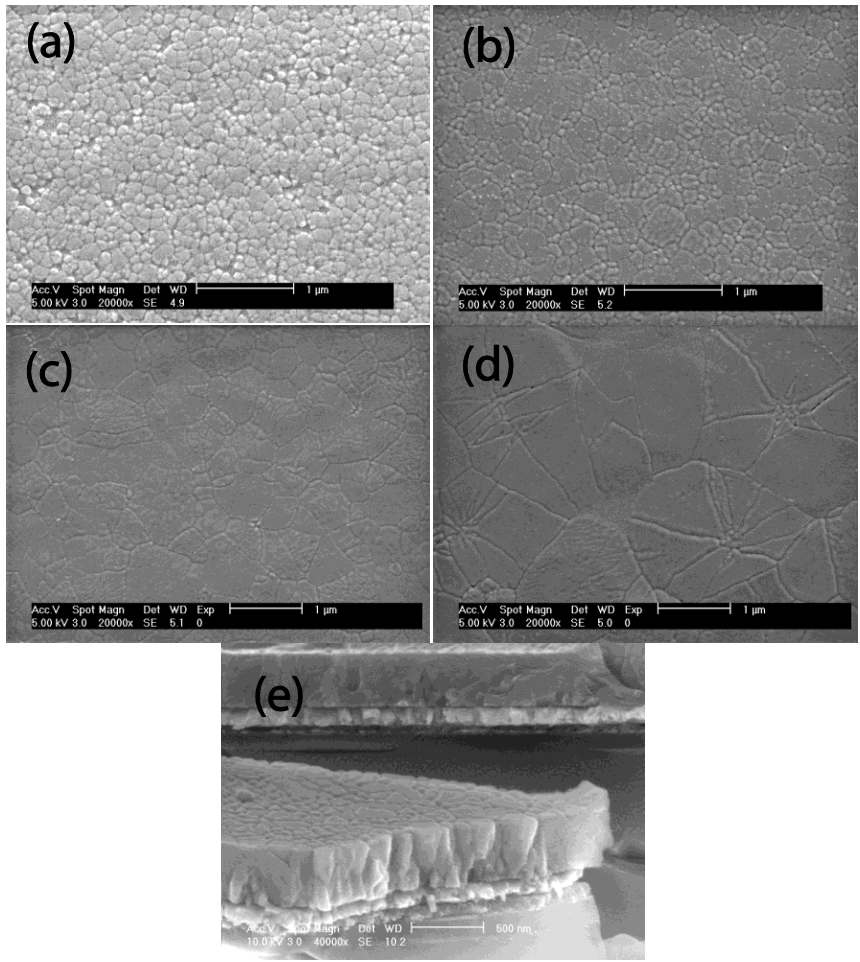

Fig 2

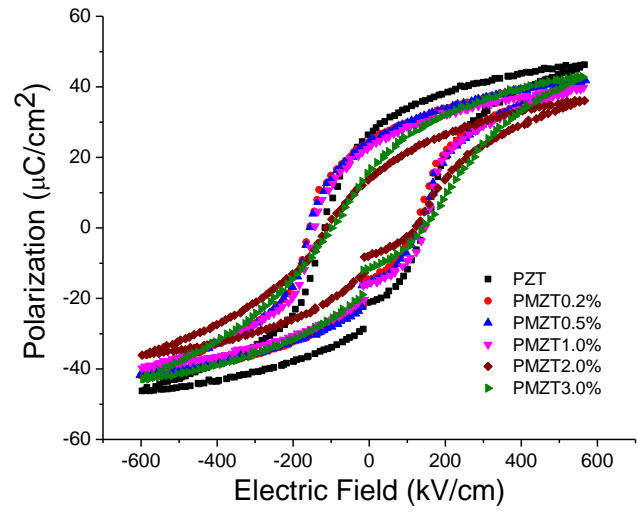

Fig 3

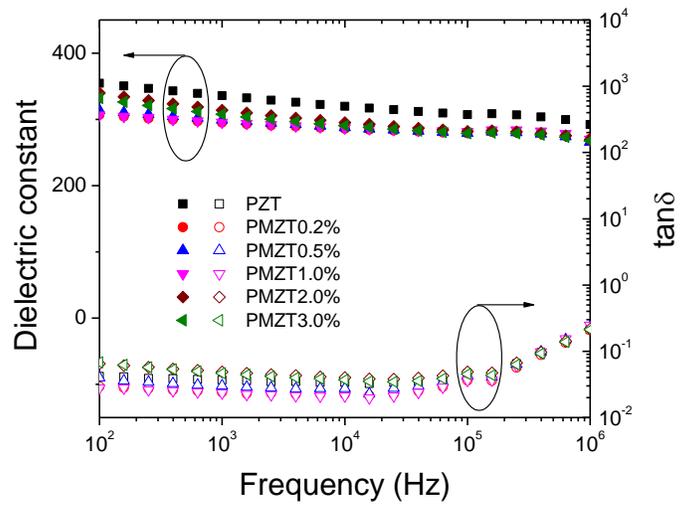

Fig 4

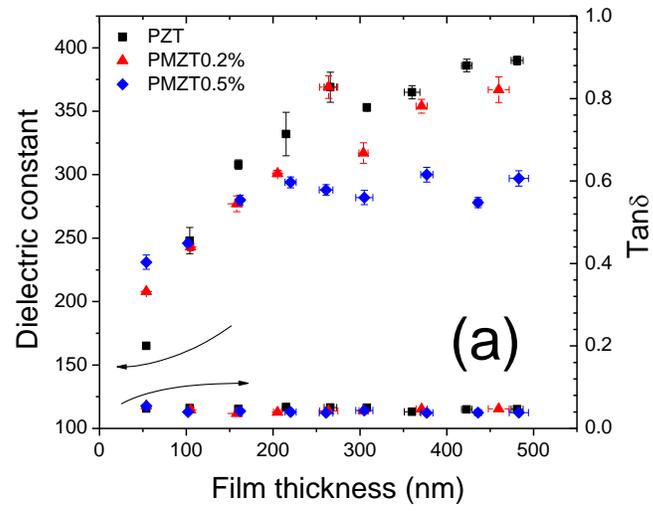

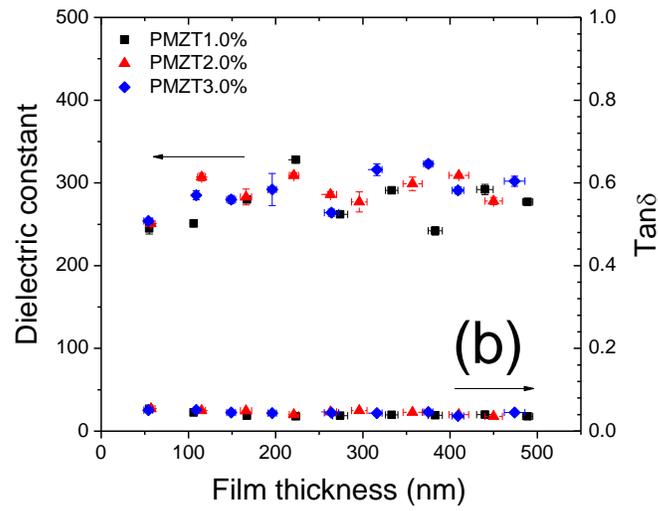

Fig 5

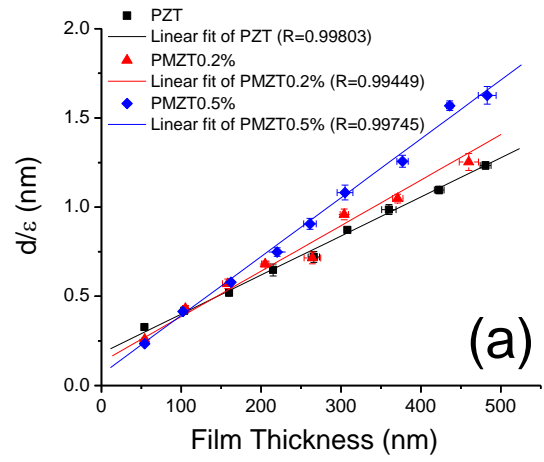
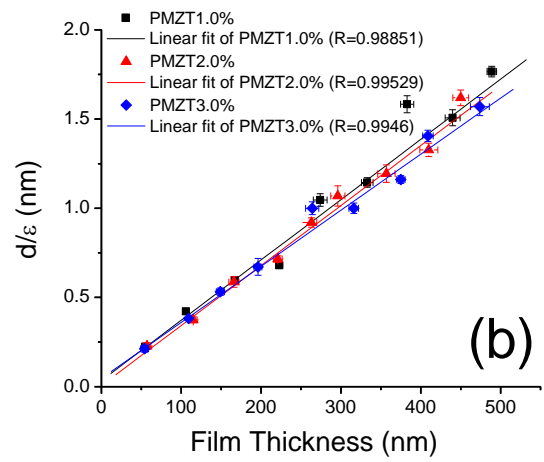
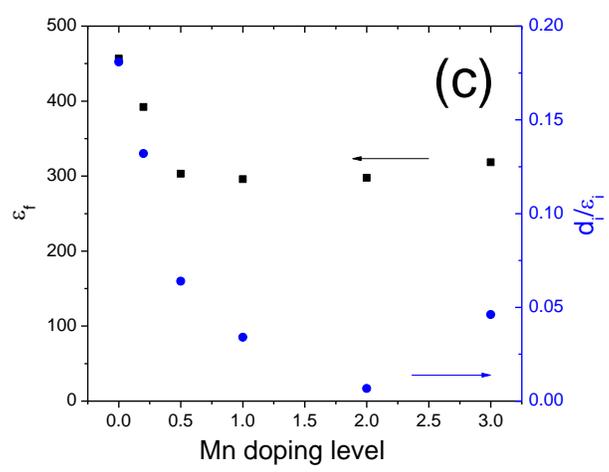

Fig 6